# Collisional energy transfer in ethanimine + He system

Vivek Vijay,[*] Francesca Tonolo,[*] Ernesto Quintas-Sánchez,[*] Adrian Batista Planas,
Carolin Joy, Richard Dawes, François Lique[†] and Dmitri Babikov[‡]

**Abstract:** The ethanimine molecule, $CH_3CHNH$, is one of the prebiotic molecules detected by astronomers in chemically-rich molecular clouds in the Galactic Center. The observations indicate a non-equilibrium distribution of rotational state populations in both the *E*- and *Z*-isomers of ethanimine, resulting from the competition between radiative processes and collisions with background gases such as He and $H_2$. Accurate interpretation of these observations requires the use of radiative transfer models with collisional state-to-state transition processes included. Here, in order to compute cross sections for state-to-state transitions in both ethanimine isomers, accurate potential energy surfaces for their interaction with a He atom were constructed and three complimentary methods for inelastic scattering were utilized: full-quantum coupled-channel and coupled-states methods, and the mixed quantum/classical theory. Strong propensities of transitions toward $\Delta k_a = 0$ and either $\Delta j = \pm 1$ (with $\Delta k_c = 0, \pm 1, \pm 2$) or $\Delta j = \pm 2$ (with $\Delta k_c = \pm 2$) are reported and the origin of this effect is identified. Small but non-negligible differences between energy transfer in the two isomers, on the order of 10%, were found. The utility of the mixed quantum/classical approach to collisional energy transfer at higher collision energies is discussed.

Author affiliations:

VV, CJ, DB -- Marquette University, Milwaukee, Wisconsin, USA

FT, FL – University of Rennes, Britany, France

EQS, ABP, RD – Missouri University of Science and Technology, Rolla, MO, USA

---

[*] These authors contributed equally

[†] Corresponding author: Francois.Lique@univ-rennes.fr

[‡] Corresponding author: Dmitri.Babikov@marquette.edu

## I. Introduction

Ethanimine ($CH_3CHNH$) is one of the prebiotic molecules detected by astronomers in chemically-rich and relatively dense molecular clouds in the Galactic Center.[1–4] In such interstellar clouds, the collisions of target molecules that emit light (such as ethanimine) with background gases (such as $H_2$ and He) compete with light emission/absorption processes. Quite often, the density of background gases is not high enough to bring the populations of rotational states of the target molecules to the equilibrium Boltzmann distribution, which creates a non-equilibrium distribution of states specific to the physical conditions of the environment. Therefore, accurate interpretation of observed spectra requires detailed modeling of radiation transfer without the assumption of local thermodynamic equilibrium (LTE) which, in turn, requires as input a set of rate coefficients for transitions between the individual rotational states of the target molecule due to collisions with background gases. This information is missing for ethanimine, forcing astronomers to use oversimplified models[5] that are rarely accurate and are very likely to lead to misinterpretation of astrophysical observations.

In this paper, we report results of the first focused study of collisional energy transfer in ethanimine molecules. For this purpose, using the tools of electronic structure theory,[6] we constructed two accurate potential energy surfaces (PESs) that represent the interaction of *E*- and *Z*-isomers of ethanimine with He atoms. Using these PESs, we carried out calculations of cross sections for rotational state-to-state transitions in ethanimine using several complimentary approaches. Namely, the quantum scattering close-coupling (CC) method[7] was used to describe $CH_3CHNH$ + He collisions at lower energies, while an approximate coupled-states (CS) method[8] was used at higher collision energies. In addition, the mixed quantum/classical theory (MQCT)[9,10] was used throughout the entire range of energies. These calculations reveal several clear patterns (propensities) of collisional energy transfer through the rotational states of ethanimine and indicate noticeable differences between its *E*- and *Z*-isomers.

The paper is composed as follows: The details of the electronic structure calculations and the construction of fitted PESs are presented in Sec. II. All details of the scattering calculations, both full-quantum and MQCT, are reported in Sec. III. Major trends of collisional energy transfer are discussed in Sec. IV. Conclusions and future directions are summarized in Sec. V.

## II. Potential Energy Surfaces

Two potential energy surfaces were constructed for this study, representing each isomer of ethanimine interacting with helium. The geometries for both isomers (*E* and *Z*) were held rigid using previously reported vibrationally averaged geometric parameters, which are consistent with the observed rotational constants.[11] With each ethanimine molecule oriented in its principal axis frame (which includes having its center of mass at the origin, and the Cs molecular symmetry plane in the x-y plane), interactions with a helium atom are described by three intermolecular coordinates: $R$, $\theta$, and $\varphi$. $R$ represents the distance between the center of mass of the ethanimine molecule and the helium atom, while $\theta$ and $\varphi$ represent the spherical angles, as illustrated in Figure 1 for the *E*-isomer. Exploiting the systems' plane of symmetry, energies were only computed in the reduced angular range: $0 < \theta < \pi/2$, and $0 < \varphi < 2\pi$. All ab initio calculations were performed using the Molpro electronic structure code package.[12]

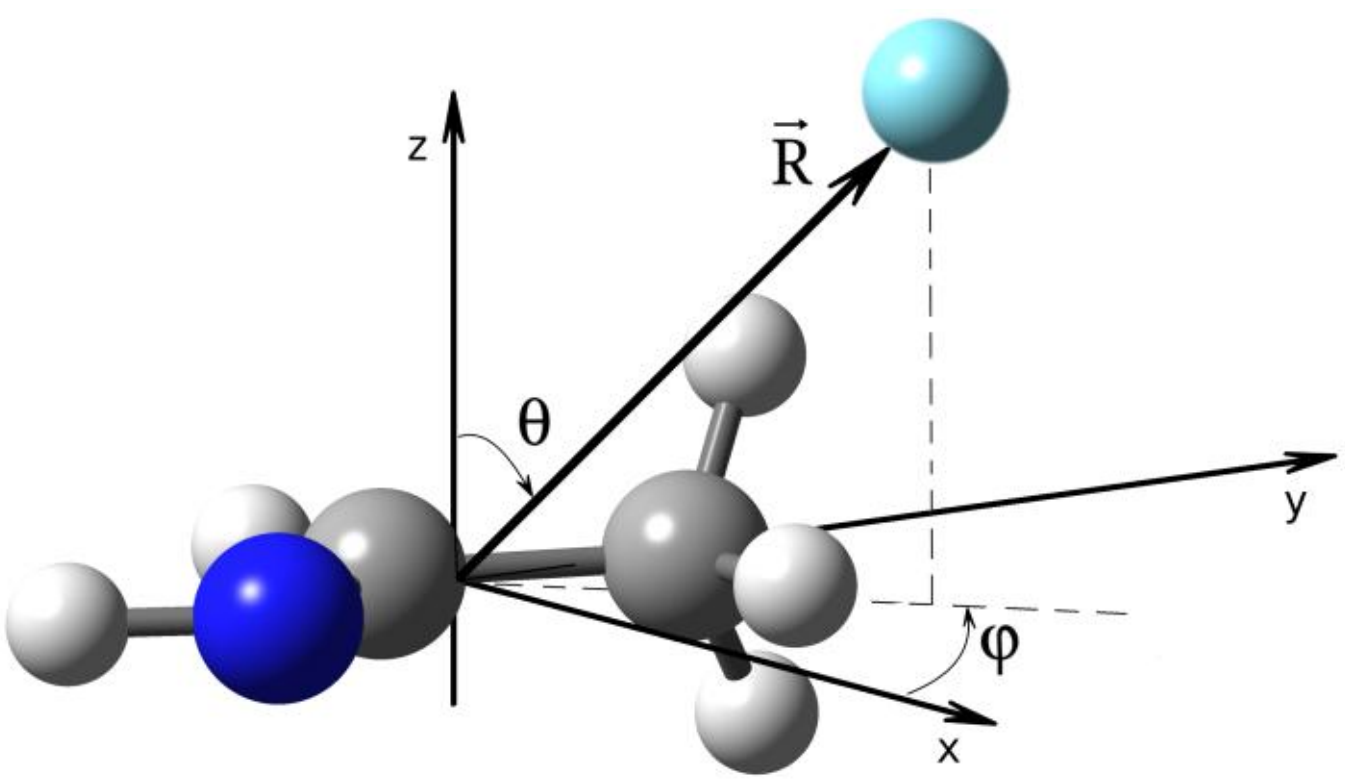

**Figure 1:** The coordinates used to describe the interaction of the ethanimine molecular isomers (*E* and *Z*, *E*-isomer shown here) with the helium atom are illustrated. See the text for details.

The non-bonded interaction energies were computed using explicitly correlated coupled-cluster theory [CCSD(T)-F12b/VTZ-F12]. As has been discussed in previous studies, the helium atom presents a slight complication because corresponding default auxiliary basis sets needed for the F12 calculations have not been published, and some irregularities have been noted in the long range interactions with helium for some systems that have been attributed to issues with the auxiliary bases.[7,13] Here, as we have done in the past for systems involving helium, the augmented correlation-consistent valence quintuple-ζ basis from the second-order Møller−Plesset perturbation theory (av5z/mp2fit) and segmented contracted highly polarized quadruple-ζ valence quality (def2-qzvpp/jkfit) basis sets were specified for the density fitting (DF) and resolution of the identity (RI) bases, respectively. No convergence issues or irregularities were detected within the fitted range 2.3 < $R$ < 18 Å. To avoid placing expensive high-level data in energetically inaccessible regions, a lower-level guide surface was first constructed. This was done using data at the MP2 level of theory which is sufficient to survey the onset of repulsion.

As we have done in the past for other three-dimensional systems,[7,13,14] the interaction potential was represented analytically with the interpolating moving least squares (IMLS) methodology and was constructed using AUTOSURF software.[6] As usual,[15,16] a local fit was expanded about each data point, and the final potential is obtained as the normalized weighted sum of the local fits. The fitting basis and most other aspects of the IMLS procedure were the same as for other previous systems and have been described in detail elsewhere.[6,15,16] This interpolative approach can accommodate arbitrary energy-surface topographies, being particularly advantageous in systems such as these, with large anisotropy.

For both ethanimine isomers, the shortest intermonomer center-of-mass distance considered is $R$ = 2.3 Å, with the additional restriction of a maximum repulsive energy of 8 kcal/mol (~ 2800 cm$^{-1}$) above the separated monomers' asymptotes. The ab initio data coverage in the fitted PESs extends to $R$ = 18 Å, which is smoothly extended by an analytic expression representing induction and dispersion interactions between the fragments (truncated at eighth order), with the zero of energy set at infinite separation between the monomers. The long-range representation was produced using the recently released LRF software.[17] For the $Z$-isomer, the global root-mean-squared (rms) fitting error in the final PES is 1.2 cm$^{-1}$, and the total number of automatically generated symmetry-unique points needed to reach that target was 2392 (the final estimated error is 0.07 cm$^{-1}$ for energies below the asymptote); while for the $E$-isomer, the global rms is 0.96 cm$^{-1}$ (0.06 cm$^{-1}$ for energies below the asymptote), with 2492 ab initio points. To guide the placement of high-level data, a lower-level guide surface was constructed using 3574 symmetry-unique points for the $Z$-isomer (3232 points for the $E$-isomer), distributed using a Sobol sequence biased to sample the short-range region more densely. The analytical representations of the PESs are available from the authors upon request.

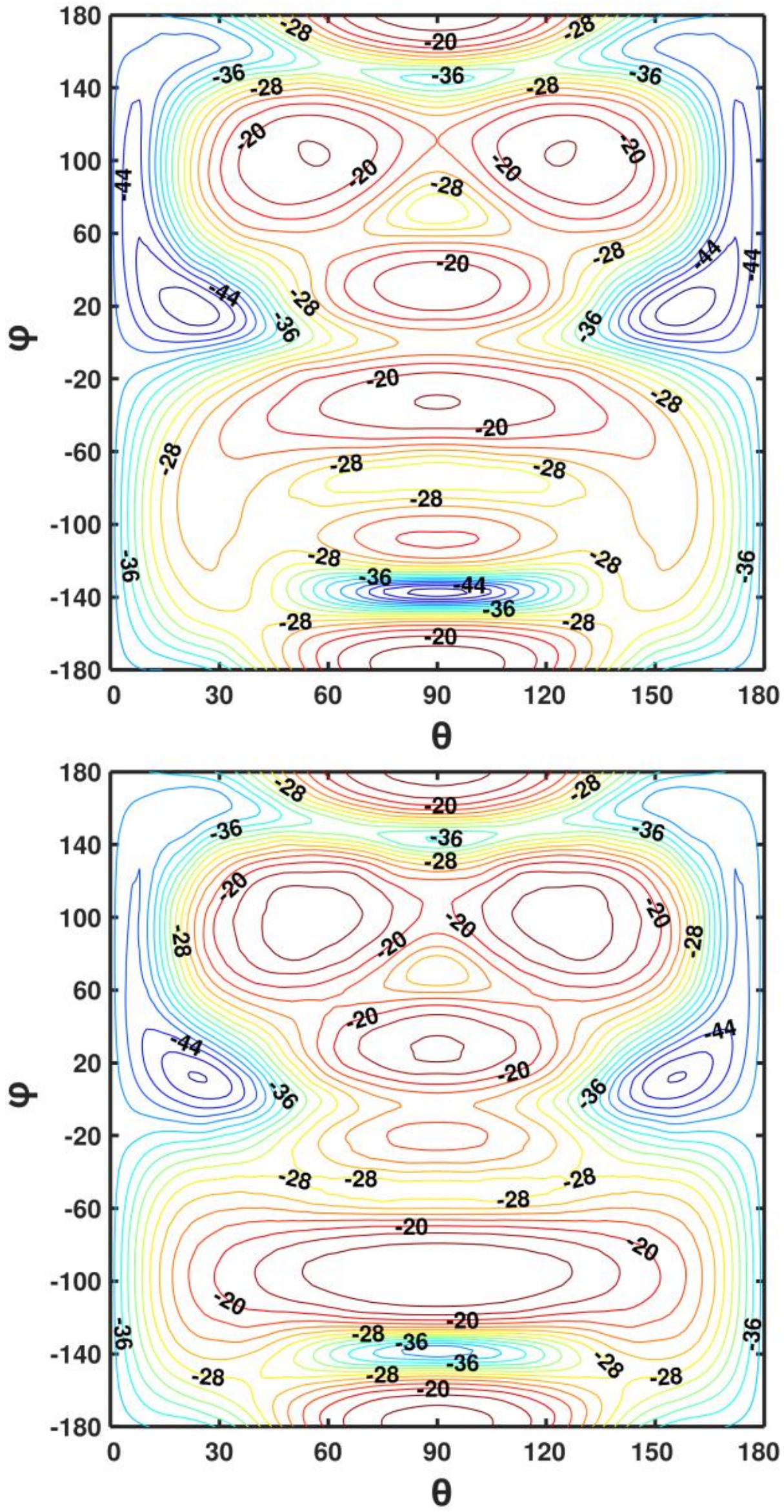

**Figure 2:** *R*-optimized contour plot of the PES for each isomer (upper: *E*-ethanimine + He, lower: *Z*-ethanimine + He) as a function of the spherical angles $\theta$ and $\varphi$. For each pair of angles, the energy (given in $cm^{-1}$) is optimized with respect to the center-of-mass distance *R*. See the text for details.

Figure 2 shows a 2D representation of the PES (denoted *R*-optimized) for each isomer, as a function of the angles $\theta$ and $\varphi$. The plots describe the complete angular ranges, relaxing the energy along the intermonomer distance coordinate *R* for each pair of angles. As can be seen in the figure, both PESs have qualitatively the same topography, characterized by five symmetry-unique minima. The energies and geometric parameters of the minima are given in Table 1. The global minimum (Min. 1) for each system is out-of-plane and has a symmetry equivalent partner. The other four minima for each system are in the mirror plan and therefore unique. Given the different structures of the *E*- and *Z*-isomers, it's remarkable how similar the sets of coordinates and well depths are. By a small margin, *Z*-ethanimine has the deeper global minimum (51.7 vs 49.5 $cm^{-1}$), associated with the closest interaction distance (3.275 Å) of any of the 10 combined minima of the two systems.

**Table 1**. Geometric parameters and well depths for the four minima of the helium atom + molecule complex. Distances are in Angstroms, angles in degrees, and energies in $cm^{-1}$.

| Ethanimine *E* + He | | | | |
|---|---|---|---|---|
| | $R$ | $\theta$ | $\varphi$ | $V$ |
| Min. 1 | 3.311 | 22.8 | 18.5 | -49.5 |
| Min. 2 | 3.936 | 90.0 | 144.7 | -36.8 |
| Min. 3 | 4.598 | 90.0 | 73.1 | -31.9 |
| Min. 4 | 4.533 | 90.0 | -73.0 | -31.7 |
| Min. 5 | 3.873 | 90.0 | -137.5 | -49.3 |
| **Ethanimine *Z* + He** | | | | |
| | $R$ | $\theta$ | $\varphi$ | $V$ |
| Min. 1 | 3.275 | 23.0 | 11.8 | -51.7 |
| Min. 2 | 3.889 | 90.0 | 143.0 | -39.7 |
| Min. 3 | 4.622 | 90.0 | 70.7 | -31.2 |
| Min. 4 | 4.342 | 90.0 | -50.4 | -32.3 |
| Min. 5 | 3.864 | 90.0 | -138.4 | -45.1 |

## III. Scattering calculations and prediction of rate coefficients

For scattering calculations using MQCT, each PES was expressed through Euler angles $(\alpha, \beta, \gamma)$ and expanded over a set of analytic functions $\tau_{\lambda\mu}$ with coefficients $v_{\lambda\mu}$:

$$V(R, \alpha, \beta, \gamma) = \sum_{\lambda\mu} v_{\lambda\mu}(R)\, \tau_{\lambda\mu}(\alpha = 0, \beta, \gamma) \tag{1}$$

where

$$\tau_{\lambda\mu}(\alpha = 0, \beta, \gamma) = \frac{1}{1+\delta_{\mu 0}}\left[(-1)^{\mu} \mathrm{Y}_{\lambda}^{+\mu}(\beta, \gamma) \; + \; \mathrm{Y}_{\lambda}^{-\mu}(\beta, \gamma)\right] \tag{2}$$

with $\lambda \geq 0$ and both even and odd values of $\mu$ included ($0 \leq \mu \leq \lambda$). In the full-quantum scattering calculations, for both CS and CS methods, an expansion equivalent to this one is employed but using spherical coordinates for $\tau_{\lambda\mu}(\theta, \varphi)$ as implemented in the MOLSCAT scattering code, which, however, is expected to give the same radial dependence of expansion coefficients $v_{\lambda\mu}(R)$. Therefore, the expansion coefficients $v_{\lambda\mu}(R)$ were computed using both MQCT code and the VRTP functionality of MOLSCAT code, on a grid of 100 points spaced logarithmically through the range $4.5 \leq R \leq 30\ a_0$. We found that the two codes produced equivalent results for $v_{\lambda\mu}(R)$, besides negligibly small numerical differences. It should be noted that this PES expansion requires that the plane of symmetry of the molecule (or, in the case of MQCT code, the reference orientation for Euler rotations) lays in the x-z plane. Therefore, prior to the expansion of the PES, the potential was transformed to place ethanimine molecule into the x-z plane such that its three rotational constants relative to x, y and z axes are 1.769, 0.289 and 0.324 $\mathrm{cm}^{-1}$ (in this order). The code written for this transformation is available from authors upon reasonable request.

The expansion was truncated to 138 terms, including all $\mu$ values for all $\lambda$ up to $\lambda = 15$, plus two more $\lambda = 16$ terms with $\mu = 0$ and 1. As a test of truncation, we also considered two larger expansions that contained 185 and 266 terms (up to $\lambda = 18$ and $\lambda = 22$, respectively, with $\mu \leq 18$ in both cases). We found, by preliminary MQCT calculations, that these extra expansion terms had a negligible effect on large cross sections of intense transitions, and only a small effect (less than 2% at $U = 100\ \mathrm{cm}^{-1}$ and below 10% at $U = 1000\ \mathrm{cm}^{-1}$) on cross sections of weak transitions. Based on this, the smaller expansion was adopted for all our calculations since it allowed us to save a significant amount of CPU time and memory for the scattering calculations.

In Fig. 3, we present the radial dependence $v_{\lambda\mu}(R)$ for 28 expansion terms up to $(\lambda, \mu) = (6,6)$ for the case of the *E*-isomer + He. One can notice that in addition to the isotropic term $(\lambda, \mu) = (0,0)$, two quadrupole terms $(\lambda, \mu) = (2,0)$ and $(2,2)$ dominate the expansion at both small and large values of distance $R$, with one more dipole term $(\lambda, \mu) = (1,1)$ weaker at large values of $R$, but becoming quite important at small $R$. All other expansion coefficients are noticeably smaller, and this is true for both isomers of ethanimine. A comparison of these most important expansion coefficients for the *E*- and *Z*-isomers is presented in Fig. S1 of Supplemental Information (SI). From Fig. S1 it follows that even anisotropic terms ($\lambda = 2$ terms, red curves in Fig. 3) are slightly larger in the *E*-isomer, while odd anisotropic terms ($\lambda = 1$ terms, green curve in Fig. 3) are slightly larger in the *Z*-isomer.

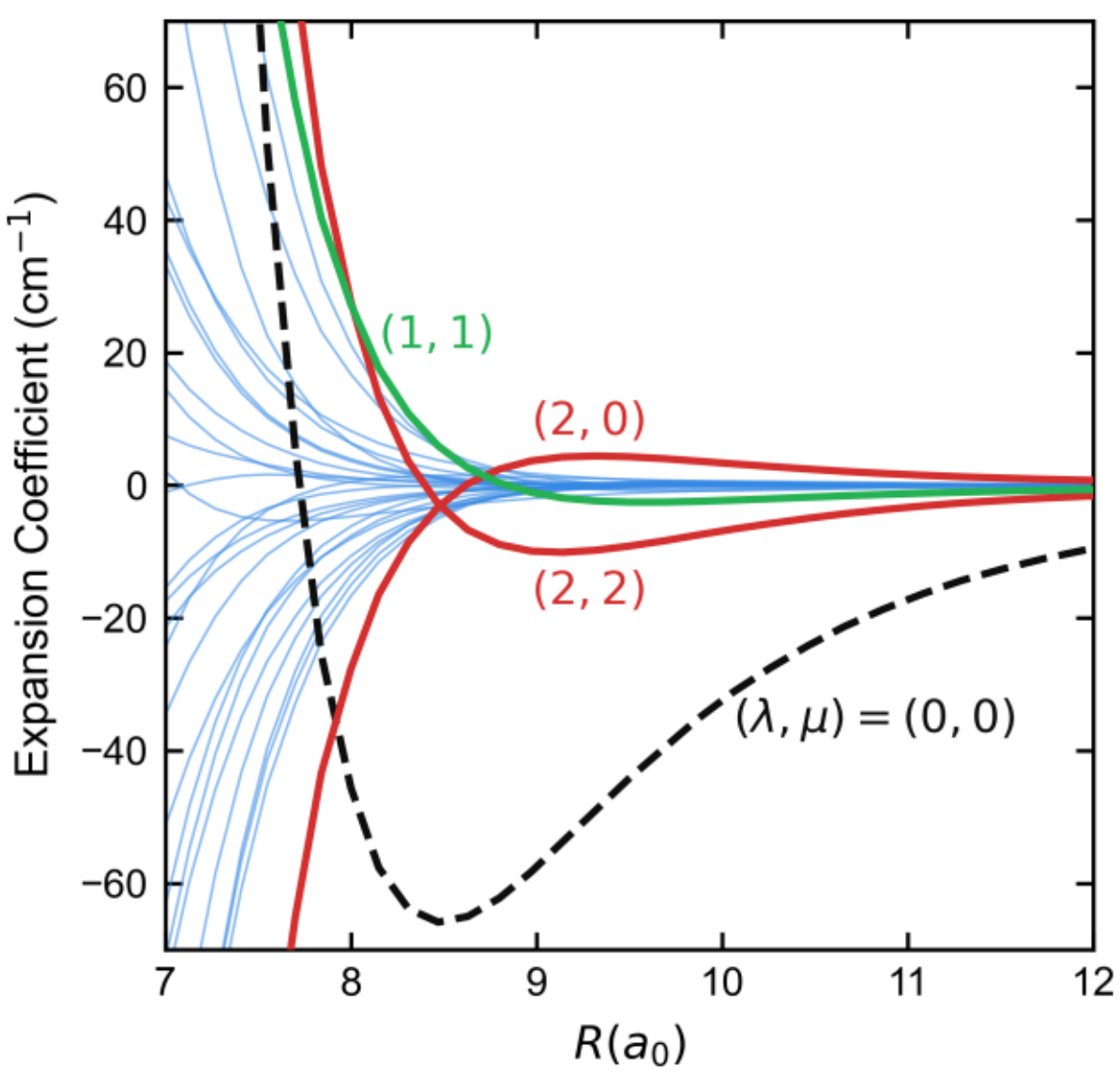

**Figure 3:** Radial dependence of 28 expansion coefficients $v_{\lambda\mu}(R)$ up to $(\lambda, \mu) = (6,6)$ for the *E*-isomer of ethanimine interacting with a He atom. Four dominant expansion terms are labeled in the figure and are color-coded for clarity.

At this point it is worth highlighting similarities and differences between the full-quantum (CC and CS) and mixed quantum/classical theory (MQCT) methods for calculations of inelastic scattering. The essential difference between these methods is in the description of translational motion of the two collision partners. In the CC method, the scattering is treated most rigorously using quantum mechanics and including the Coriolis coupling effects, while in the CS method the Coriolis coupling is approximated (an assumption justified at higher collision energies). In the MQCT method the Coriolis coupling is included, but the scattering is described classically using mean-field trajectories, which is also an approximation that may become inaccurate at low collision energy. However, all three methods use the same expansion of the PES, such as Eqs. (1-2). Also, all three methods use the same set of asymmetric-top-rotor eigenstates, *i.e.*, wavefunctions $\Psi_i(\alpha,\beta,\gamma)$ and energies $E_i$, to describe rotation of the target molecule (here ethanimine). Therefore, the matrix of potential coupling between these rotational states, $M_{i\to f}(R) = \langle \Psi_f(\alpha,\beta,\gamma) | V(R,\alpha,\beta,\gamma) | \Psi_i(\alpha,\beta,\gamma) \rangle$, is also the same in MQCT and in the full quantum methods formulated in the body fixed-reference frame, such as the CS method implemented in MOLSCAT and HIBRIDON programs.[18,19] Due to these similarities, different methods employed in this work have a common set of input parameters, as described next.

For MQCT and CS calculations the rotational basis set used for ethanimine contains all rotational states up to $E_{\max} = 140\ \mathrm{cm}^{-1}$, which includes 237 states up to $j_{k_a k_c} = 21_{1,21}$ for the *E*-isomer, and 245 states up to $j_{k_a k_c} = 18_{5,13}$ for the *Z*-isomer. Preliminary MQCT and CS calculations for $U = 1000\ \mathrm{cm}^{-1}$ showed that, with this basis set, cross sections for transitions between the lowest 86 states (up to $j_{k_a k_c} = 6_{5,1}$ at $E_i = 50\ \mathrm{cm}^{-1}$) are converged to within 2% for more intense transitions and to within 15% for less intense transitions. For CC calculations, the rotational basis set was slightly reduced to exclude states with $j > 14$ for collisions with total energy < $20\ \mathrm{cm}^{-1}$, and exclude states with $j > 16$ for collisions with total energy > $20\ \mathrm{cm}^{-1}$. It was found that this truncation gives cross sections converged to within 5%. Overall, these calculations give us reasonably well-converged cross sections for 3655 transitions between the individual rotational states of the *E*-isomer, and for 3828 transitions of the *Z*-isomer.

MQCT calculations were carefully checked for tight convergence with respect to the maximum impact parameter $b_{\max}$ that determines the maximum value of the orbital angular momentum quantum number $\ell_{\max}$ for collision. For example, the values of $b_{\max} = 25,\ 20,$ and $15$ were used for collision energies $U = 1, 40$ and $600\ \mathrm{cm}^{-1}$, respectively, which translates into $\ell_{\max} = 6,\ 38,$ and $151$, respectively. Similar values of the total angular momentum $J_{\mathrm{tot}}$ were used in full-quantum CS and CC calculations. A test of MQCT state-to-state transition cross sections with respect to microscopic reversibility is presented in Fig. S2 of SI. For each transition, both quenching and excitation cross sections were computed, and their weighted average was used to obtain final results that satisfy microscopic reversibility, as explained in our recent papers.[20,21] Quantum state-to-state transition cross sections automatically satisfy microscopic reversibility.[22]

Figure 4 shows several typical examples of the energy dependencies of state-to-state transition cross sections computed using the three different methods (CC, CS and MQCT) for strong transitions with large cross sections. Figure S3 in SI gives several examples of similar data for weak transitions with small cross sections (the origins of strong and weak transitions are discussed in detail in the next section). From this data, we conclude that at low collision energies MQCT method underestimates cross sections by a factor of 2-3 for strong transitions. For weak transitions the difference is sometimes larger (see SI). It should be noted that at low collision energies, $E_{\mathrm{col}} < 40\ \mathrm{cm}^{-1}$, many MQCT trajectories are trapped in the interaction region (described as orbiting of two collision partners and/or vibration of the metastable complex), which is a classical analogue of quantum scattering resonances. These trajectories do not leave the interaction region and cannot be analyzed in a standard way. In this work we tried to propagate trapped trajectories for at least five periods of orbiting (or five periods of vibration, whatever happens first) but, still, the transition probabilities and cross sections from MQCT are *smaller* than those obtained from the full-quantum CC calculations. From Fig. 4 we can also see that cross sections obtained using the approximate CS-method are *larger* than those from accurate CC-calculations. In fact, we found that for many strong state-to-state transitions at low collision energy the values of accurate cross sections obtained using CC method are in between the results of SC and MQCT methods, as one can see in Fig. 4.

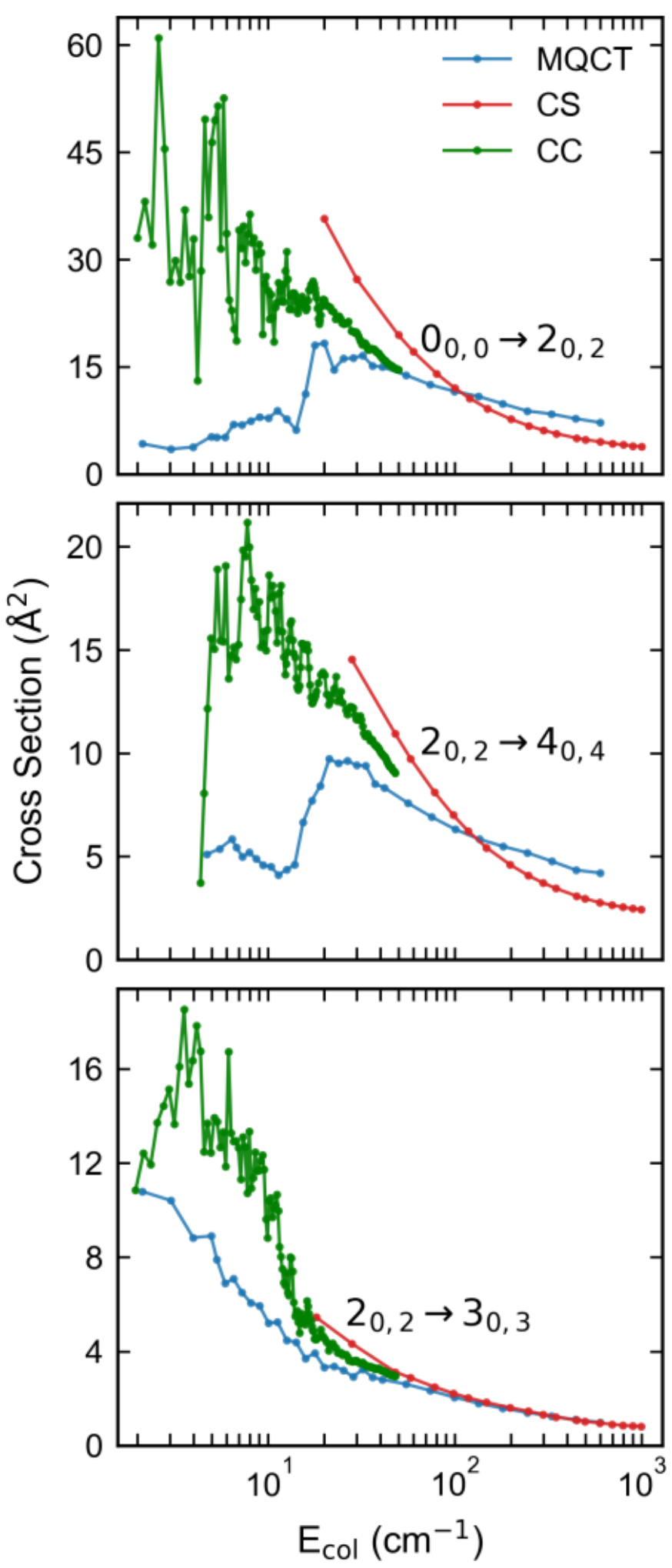

**Figure 4:** Collisional cross sections for three strong transitions between the rotational states of the *E*-isomer of ethanimine as a function of collision energy. The projectile is a He atom. The results of three methods are shown by different colors as indicated in the legend.

However, as the collision energy increases, the predictions of CS and MQCT methods become more reliable. For some transitions, like $2_{0,2} \rightarrow 2_{0,3}$ in the lower frame of Fig. 4, the results of all three methods become very similar in the range $30 < E_{\mathrm{col}} < 50\ \mathrm{cm}^{-1}$, whereas in the range of $E_{\mathrm{coll}} > 50\ \mathrm{cm}^{-1}$ they become nearly identical. We found that, for many strong transitions at intermediate collision energies, the predictions of all three methods become similar. At higher collision energies, where the results of an accurate but expensive CC method are unavailable, the predictions of CS and MQCT methods are quite similar. This is further discussed in Sec. IV.

## IV. Major trends of collisional energy transfer

The mixed quantum/classical and the full-quantum calculations of rotationally inelastic scattering in the ethanimine + He system reveal the same major trends of collisional energy transfer. In Fig. 5 we present the data computed using MQCT for 3655 state-to-state transitions in ethanimine *E*-isomer collided with a He atom at kinetic energy $U = 600\ \mathrm{cm}^{-1}$. Cross sections for excitation and de-excitation processes are plotted as a function of rotational energy change in the molecule, $\Delta E = E_f - E_i$, that can be both positive and negative. Indexes "$i$" and "$f$" are used to label the initial and final rotational states of the target molecule. Similar data obtained using the CS method of MOLSCAT for a total energy of $600\ \mathrm{cm}^{-1}$, are presented in Fig. S4 of SI.

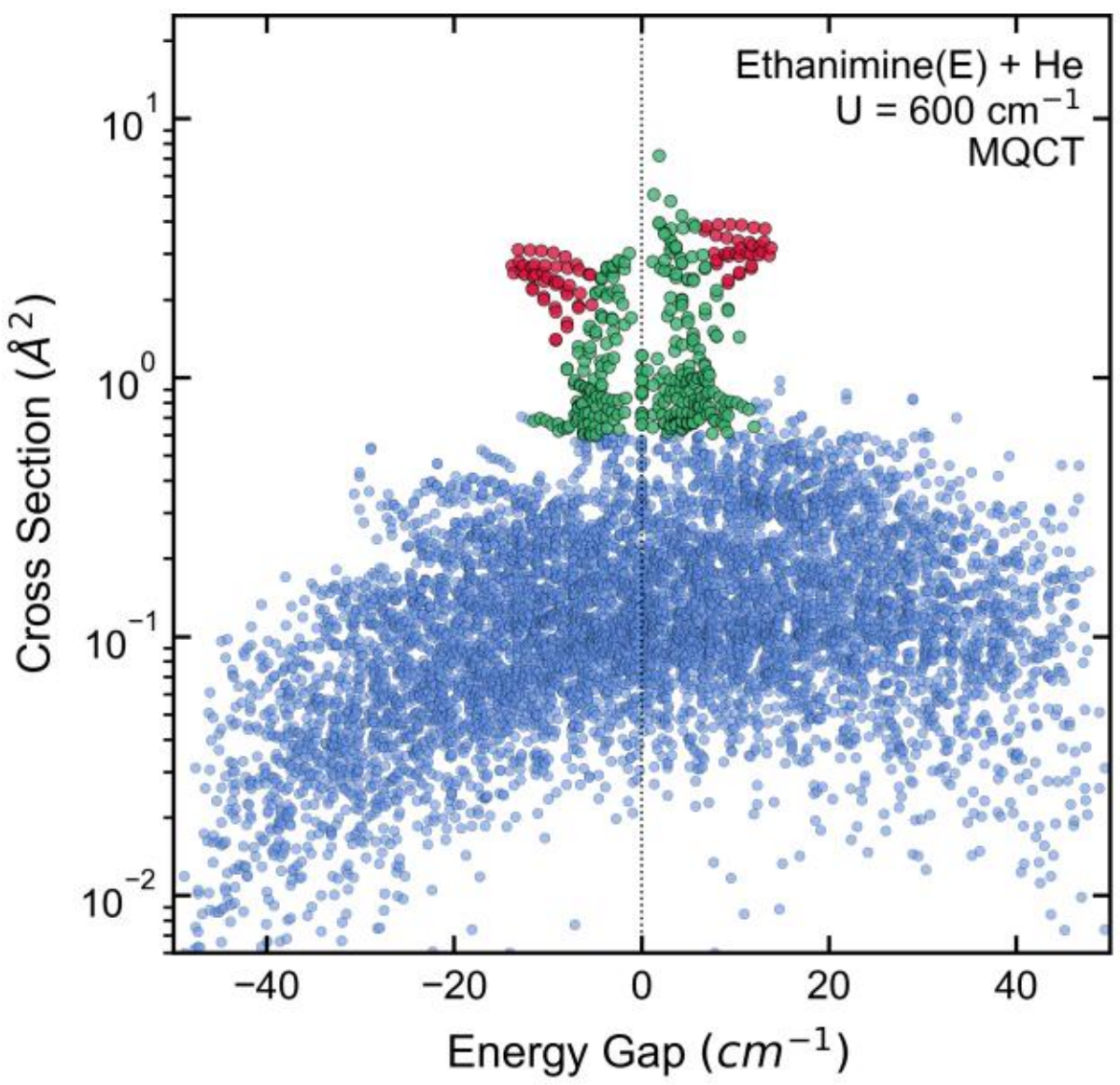

**Figure 5:** State-to-state transition cross sections computed by MQCT for rotational transitions in ethanimine $E$-isomer collided with He atom at kinetic energy $U = 600\ \text{cm}^{-1}$, as a function of collisional energy transfer $\Delta E$. Color indicates three distinct groups of transitions and is consistent with the colors of Fig. 3 as explained further in the text.

From Fig. 5 and Fig. S4 one can see that, although on a qualitative level, the trend of data is somewhat similar to the well-known exponential gap law,[23] several distinct features indicate the influence of at least three different mechanisms of energy transfer in ethanimine, as indicated by three colors in these figures. Namely, a compact group of datapoints (shown in Fig. 5 in red) that correspond to relatively large cross sections, 3 to 4 $Å^2$, and moderate energy transfer, $8 < |\Delta E| < 16\ \text{cm}^{-1}$, is distinct from another group (shown in Fig. 5 in green), where the energy transfer is small, $|\Delta E| < 8\ \text{cm}^{-1}$, but the values of the cross sections vary across more than an order of magnitude range, from 0.5 to 8 $Å^2$. Moreover, these two relatively small groups of datapoints are clearly different from the main group (blue symbols in Fig. 5) that includes 95% of all transitions characterized by small cross sections, under 0.5 $Å^2$, in a wide range of energy transfer values, up to $|\Delta E| \sim 50\ \text{cm}^{-1}$.

Importantly, the very same features are present in cross sections we obtained for the $Z$-isomer of ethanimine using both MQCT method (as presented in Fig. S5 of SI) and CS method of MOLSCAT (as presented in Fig. S6 of SI).

In order to identify the origin of the three energy transfer mechanisms in ethanimine, we processed all data from Fig. 5 by binning cross section values into a 2D-histogram shown in Fig. 6, where the two axes correspond to the change of rotational quantum numbers for transitions in ethanimine. Namely, the horizontal axis describes the change of angular momentum quantum number $\Delta j = j_f - j_i$, while the vertical axis describes the change of the projection quantum number $\Delta\tau = \tau_f - \tau_i$, where $\tau = k_a - k_c$ as usual. The area of the circles in Fig. 6 corresponds to the sum of cross sections in each bin. The color of contributions in Fig. 6 is consistent with the colors in Fig. 5.

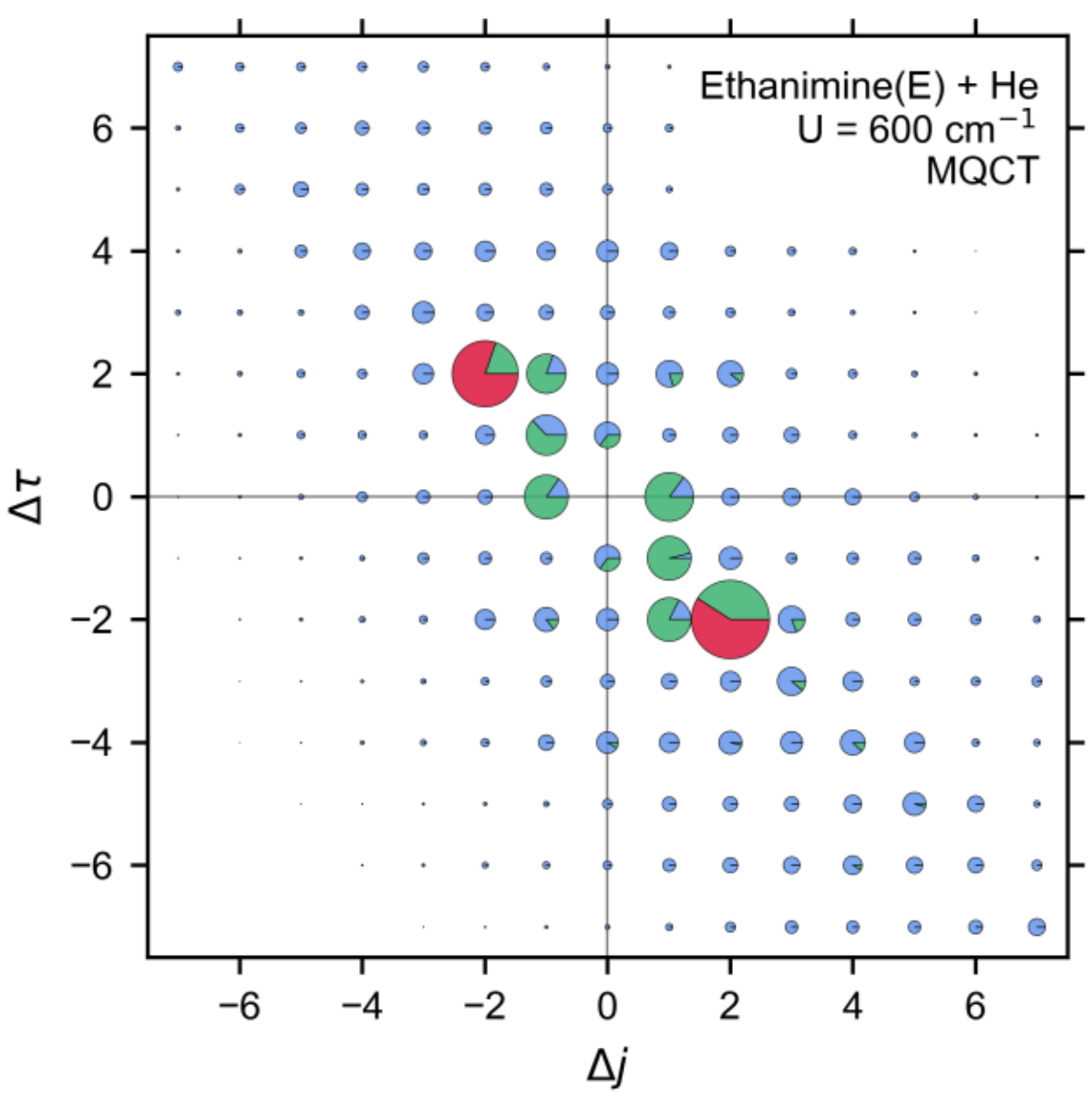

**Figure 6:** Two-dimensional histogram of the data presented in Fig. 5, to reveal the propensity rules for ethanimine. Horizontal and vertical dimensions characterize the change of rotational quantum numbers in each state-to-state transition, $\Delta j$ and $\Delta\tau$. Colors of contributions correspond to those of Fig. 5.

From this histogram one can clearly see that the compact group of datapoints shown in red in the energy transfer diagram of Fig. 5 corresponds exclusively to transitions with $\Delta j = -\Delta\tau = \pm 2$, which determines the main propensity rule for this molecule. Also, one can see that the extended group of datapoints shown in Fig. 5 in green has contributions from $\Delta j = \pm 1$ and $\Delta\tau$ with opposite sign and $|\Delta\tau| = 0$, 1 and 2, which establishes the secondary propensity rule for this molecule. Contributions of all other values of $\Delta j$ and $\Delta\tau$ are much smaller.

Importantly, one can go back to Fig. 3 and make a connection between the dominant expansion terms $(\lambda, \mu)$ and dominant cross sections in Fig. 5 (note that the color is consistent throughout these figures). Namely, it makes sense to assume that one of two dominant expansion terms $(\lambda, \mu) = (2{,}2)$ is largely responsible for the main propensity rule $\Delta j = -\Delta\tau = \pm 2$. To prove this hypothesis, we carried out additional MQCT calculations where we removed the term $(2{,}2)$ from the expansion of the PES and repeated the calculations of cross sections. As expected, we found that in this case the main propensity $\Delta j = -\Delta\tau = \pm 2$ disappears. Interestingly, the removal of another large term $(\lambda, \mu) = (2{,}0)$ leads to a very similar effect. However, when both $(2{,}0)$ and $(2{,}2)$ terms are removed, the original propensity is only partially reduced but is still present, which is likely the result of several consecutive transitions driven by another significant expansion term $(\lambda, \mu) = (1{,}1)$.

It should be pointed out that the overall effect of the secondary propensity, when summed over three values of $|\Delta\tau| = 0$, 1 and 2, is even somewhat larger than the effect of the primary propensity. Therefore, two propensity rules, with $\Delta j = \pm 1$ and $\Delta j = \pm 2$, are both important in ethanimine molecules. In terms of projection quantum numbers $k_a$ and $k_c$ they both result in a strong propensity toward transitions with $\Delta k_a = 0$ and either $\Delta k_c = \pm 2$ (for $\Delta j = \pm 2$), or $\Delta k_c = 0, \pm 1, \pm 2$ (for $\Delta j = \pm 1$). Several of the most intense transitions, in decreasing order and in the direction of quenching, are: $11_{1,10} \rightarrow 9_{1,8}$, $10_{1,9} \rightarrow 8_{1,7}$, $9_{1,8} \rightarrow 7_{1,6}$, $8_{1,7} \rightarrow 6_{1,5}$, $7_{1,6} \rightarrow 5_{1,4}$, $4_{3,2} \rightarrow 3_{3,0}$, $4_{3,1} \rightarrow 3_{3,1}$, $2_{1,1} \rightarrow 1_{1,1}$, $3_{2,1} \rightarrow 2_{2,1}$, $11_{2,9} \rightarrow 9_{2,7}$, $11_{0,11} \rightarrow 9_{0,9}$, $12_{0,12} \rightarrow 10_{0,10}$, $9_{0,9} \rightarrow 7_{0,7}$, $3_{2,2} \rightarrow 2_{2,0}$. Such simple propensity rules are more typical of smaller molecules,[24–26] but several recent studies of collisional energy transfer have reported similar propensities in rather large polyatomic molecules.[27,28]

We also explored how the energy transfer diagram shown in Fig. 5 for collision energy $U = 600$ cm$^{-1}$, evolves if the collision energy is reduced. Figure S7 in SI shows similar diagrams for collision energies $U = 10$ cm$^{-1}$ and $40$ cm$^{-1}$. One can see that reducing the energy of collision to $40$ cm$^{-1}$ does not change the behavior significantly. The groups of red, green and blue datapoints slightly shift (compared to Fig. 5) but the same features are still present. However, reducing collision energy further down to $10$ cm$^{-1}$ leads to a qualitative change -- now all three groups of datapoints merge into one single structure that looks rather uniform and resembles a simple exponential model of collisional energy transfer.[23] One possible explanation for this transformation is that at low collision energy the motion of collision partners explores only the low energy part of the PES, where all expansion terms are comparable and thus all terms affect state-to-state transition processes to a certain extent. Therefore, it is impossible to identify a small number of leading terms responsible for strong propensity rules. Moreover, from a quantum standpoint, the scattering process at low collision energy is dominated by resonances, which makes the propensity much less visible. In this sense, we can say that propensity is a high collision energy phenomenon that disappears at low energy.

It should be stressed that the diagrams for the *E*-isomer presented in Figs. 5 and 6 based on MQCT calculations are very similar to those obtained using the full-quantum method (presented in two frames of Fig. S4 of SI). Similarly, the diagrams for the *Z*-isomer presented in Fig. S5 based on MQCT calculations are very similar to those obtained using the full-quantum method presented in Fig. S6. This serves as a benchmark of the applicability of the MQCT approach and permits us to conclude that at high collision energy the mixed quantum/classical approach captures all features of collisional energy transfer in both isomers of ethanimine.

Figure 7 represents a comparison of state-to-state transition cross sections for the two isomers of ethanimine. For each transition we computed the average over the two isomer's value of cross section (plotted along the horizontal axis in Fig. 7) and a percent difference between the cross sections for the two isomers (plotted along the vertical axis in Fig. 7). In this figure one can identify a group of strong transitions with large cross sections in the range between $1$ and $3$ $\text{Å}^2$. We found that for all strong transitions of the *Z*-isomer, cross sections are larger by 10% on average compared to those in the *E*-isomer. For weaker transitions with cross sections under $1\ \text{Å}^2$ the differences between the two isomers become larger, as one can see in Fig. 7, but they do not follow a clear trend anymore. The RRMS deviation of all datapoints in Fig. 7 is about $29\%$. Figure 7 represents the results of MQCT calculations. Figure S6 in SI gives the same data but obtained from CS calculations (at total energy of $600\ \text{cm}^{-1}$) and they show a similar trend: cross sections for strong transitions are larger in the *Z*-isomer of ethanimine by $10\%$ on average. For all transitions in Fig. S6 the RRMS difference between *E*- and *Z*-isomers is about $26\%$. Note, however that RRMS stands for the *relative* RMS deviation determined mostly by many weak transitions with small cross sections that are likely to play only a minor role in collisional energy transfer.

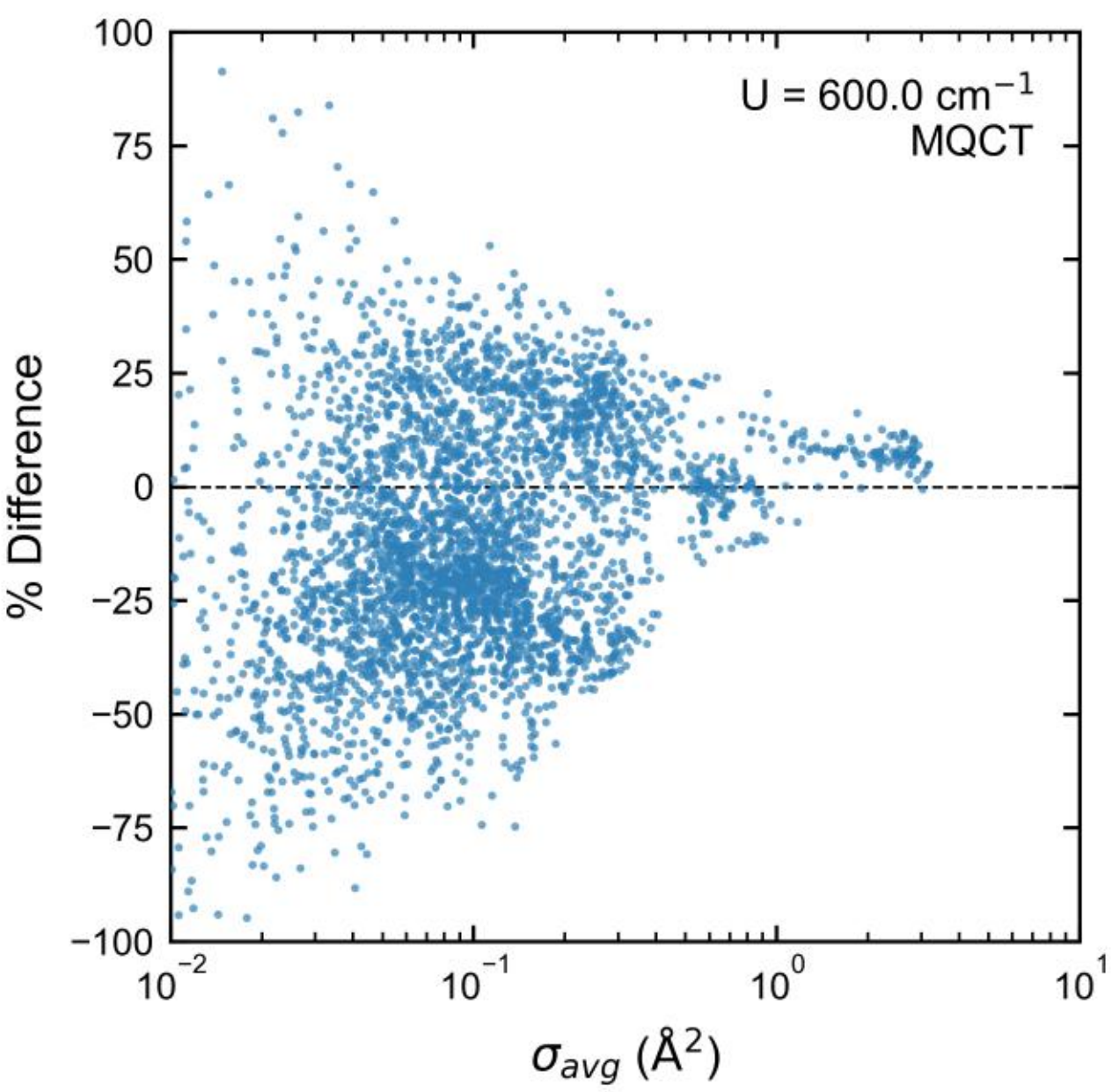

**Figure 7:** Percent difference between cross sections for state-to-state transition in two isomers of ethanimine, plotted versus the average value of cross sections for two isomers, according to MQCT calculations at collision energy $U = 600\ \text{cm}^{-1}$. For each transition only one datapoint is included, that corresponds to the quenching process.

In order to quantify the performance of MQCT across all transitions in the ethanimine + He system, we present in Fig. 8 a comparison of quenching cross sections for all 3828 rotational state-to-state transitions in both isomers, relative to the full-quantum CS calculations at a total energy of $300\ \text{cm}^{-1}$. From Fig. 8 one can see that the results of MQCT are rather accurate for more intense transitions that cover an order of magnitude range of cross-section values. For these processes MQCT just slightly overestimates cross sections. For weaker transitions, MQCT systematically underestimates the values of cross sections, but for the majority of weak transitions this difference is within a factor of two, which is considered acceptable for applications in the modeling of radiative energy transfer in astrophysical environments.

From the data obtained in this work we can project that the MQCT method can be used to compute thermal rate coefficients for rotational state-to-state transitions in the ethanimine + He system, and probably for other small polyatomic molecules, in the temperature range $T > 50$ K. At lower temperatures the effect of quantum scattering resonances becomes more important, and the full-quantum calculations at CC level should be used to provide reliable data. Blending cross sections from CC calculations at low collision energies with the MQCT cross sections at high collision energies to produce rate coefficients for a broad range of temperatures seems to represent another viable approach.

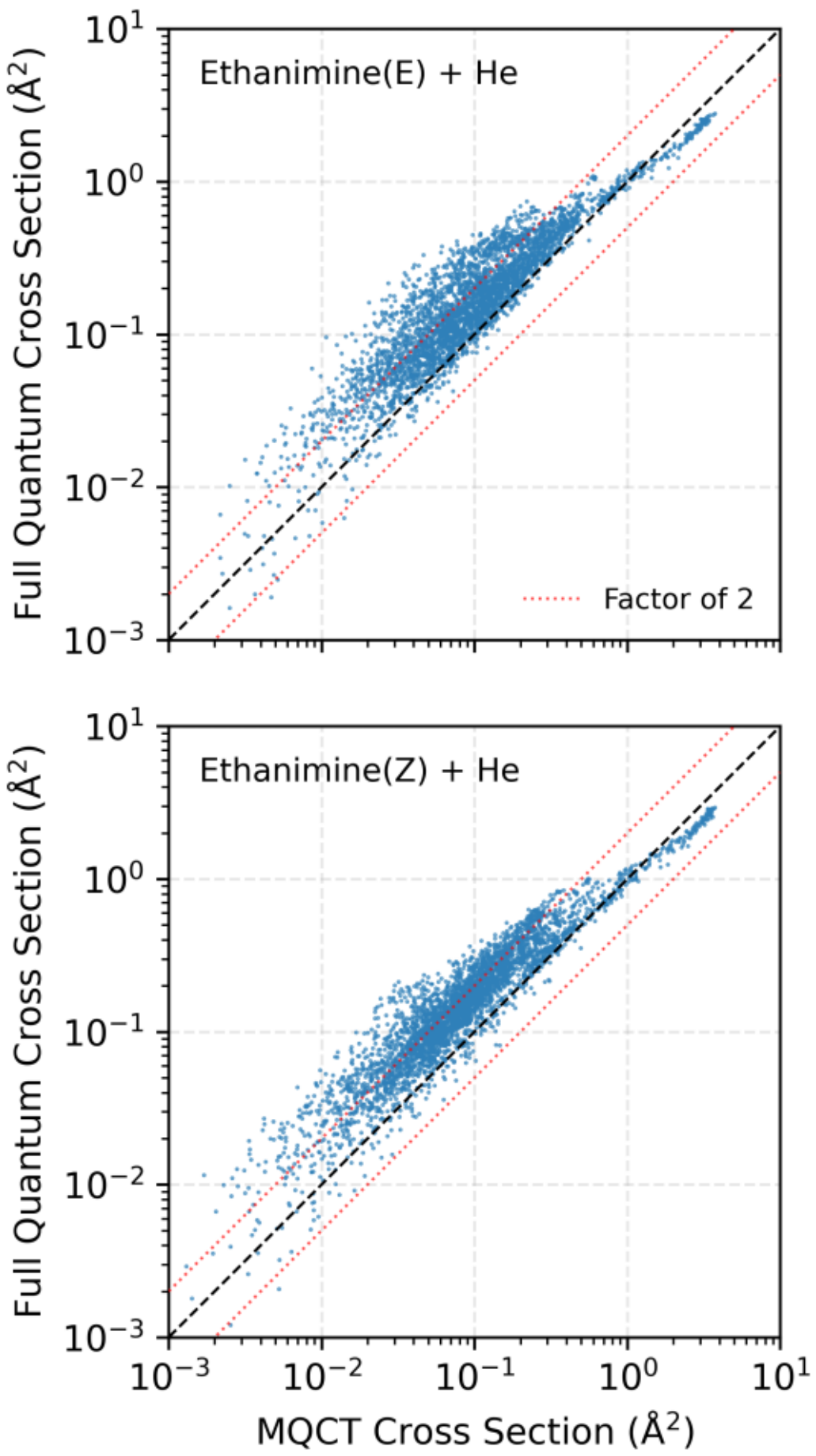

**Figure 8:** Comparison of cross sections obtained by full-quantum CS calculations with those from MQCT, for 3828 rotational state-to-state transitions for the *E*- (upper frame) and *Z*-isomers of ethanimine colliding with an He atom at total energy of 300 cm$^{-1}$.

## V. Conclusions

In this paper we conducted the first detailed study of collisional energy transfer in ethanimine, which involved the development of two accurate potential energy surfaces for the interaction of its *E*- and *Z*-isomers with an He atom and the calculations of rotationally inelastic scattering using three complimentary methods: full-quantum CC and CS methods and the mixed quantum/classical

MQCT method. The CC method gives an accurate description of quantum resonances at low collision energies (where CS and MQCT methods are likely to be less accurate) while CS and MQCT methods permit calculations at higher collision energies (where the CC method becomes computationally demanding or even not any more affordable). We found that despite the fact that from an astrochemical perspective, ethanimine is a rather complex organic molecule, its interaction with He atom is dominated by just the two largest terms of the PES expansion $(\lambda, \mu) = (2{,}0)$ and $(2{,}2)$, with a somewhat smaller contribution of the term $(1{,}1)$ and very small contributions from all other terms (out of 138 terms considered). Scattering calculations carried out using this expansion showed that the properties of ethanimine + He interaction manifest during inelastic collisions as the primary propensity towards transitions characterized by $\Delta j = \pm 2$ with $\Delta\tau = -\Delta j$, and the secondary propensity characterized by $\Delta j = \pm 1$ with $\Delta\tau$ of opposite sign and absolute values $|\Delta\tau| = 0$, 1 and 2. Importantly, these propensities create a small group of strong transitions with cross sections and rate coefficients that are an order of magnitude larger than all other state-to-state transitions in this molecule. In terms of projection quantum numbers, they are characterized by $\Delta k_a = 0$ and either $\Delta j = \pm 1$ (with $\Delta k_c = 0, \pm 1, \pm 2$) or $\Delta j = \pm 2$ (with $\Delta k_c = \pm 2$).

At this point, we can hypothesize that under conditions of cold molecular clouds such transitions may populate the excited rotational states of ethanimine selectively, pumping up only the manifold of $k_a = 0$ states (such as $j_{0,j}$) and creating a non-equilibrium distribution of rotational levels. For example, if one starts at the ground rotational state $0_{0,0}$, the two strongest collision induced transitions are $0_{0,0} \rightarrow 1_{0,1}$ and $0_{0,0} \rightarrow 2_{0,2}$. In turn, from $1_{0,1}$ the two strongest transitions are $1_{0,1} \rightarrow 2_{0,2}$ and $1_{0,1} \rightarrow 3_{0,3}$ while from $2_{0,2}$ the two strongest transitions are $2_{0,2} \rightarrow 3_{0,3}$ and $2_{0,2} \rightarrow 4_{0,4}$. Then, from state $3_{0,3}$ the two strongest are $3_{0,3} \rightarrow 4_{0,4}$ and $3_{0,3} \rightarrow 5_{0,5}$ transitions, etc. In fact, we found that as the quantum number $j$ increases, the intensity of such transitions also increases, with maximum values around $j = 9$.

An accurate description of this scenario requires rigorous modeling of radiative transfer in non-LTE conditions with collisional energy transfer included which in turn requires a full set of reliable state-to-state transition rate coefficients for the relevant range of temperatures (including very low temperatures) and with He atoms and $H_2$ molecules as projectiles. Thus, our plan for future research is to extend this work to the ethanimine + $H_2$ system and create a full database of rate coefficients for collisions of the *E*- and *Z*-isomers of ethanimine with both He and $H_2$, which will enable applications in astrophysical modelling.

Here we found that the difference between the two isomers of ethanimine is small but is non negligible. For those strong transitions that are likely to dominate the energy transfer process, cross sections and rate coefficients for the *Z*-isomer are systematically larger than those of the *E*-isomer, by about 10% on average. This property was confirmed by both full-quantum and MQCT calculations. Thus, in the future, both isomers should be considered to determine the magnitude of this difference in the case of the $H_2$ projectile, which is more abundant than He atoms and thus is more important for the modeling of astrophysical environments.

Another important conclusion of this work concerns the benchmarking of the approximate mixed quantum/classical theory approach, MQCT. It is shown that although at low collision energies, dominated by quantum scattering resonances, cross sections predicted by MQCT are smaller compared to the full quantum CC results (often within a factor of two, but in many cases by more than that) these differences are reduced as the collision energy is raised. In the high-temperature regime MQCT is expected to give a reasonable description of the energy transfer process. For strong transitions in ethanimine + He, the rate coefficients predicted by MQCT are expected to become sufficiently accurate at $T \sim 50$ K and above. In the future, this method can be used for prediction of rate coefficients for the modeling of high-temperature astrophysical environments.

**CONFLICTS OF INTEREST**

Authors declare no conflict of interests.

**DATA AVALABILITY**

All data reported in this work is available from the authors upon request.

**ACKNOWLEDGEMENTS**

This research was supported by NSF grant number CHE-2102465. This work used Anvil CPU resources at Purdue University through allocation PHY260027 from the Advanced Cyberinfrastructure Coordination Ecosystem: Services & Support (ACCESS) program, which is supported by National Science Foundation grants #2138259, 2138286, 2138307, 2137603 and 2138296. We also used resources of the National Energy Research Scientific Computing Center, supported by the Office of Science of the U.S. DoE under Contract No. DE-AC02-5CH11231. DB acknowledges the support of Way Klingler Sabbatical Fellowship. FL and FT acknowledge the Regional Council of Brittany for supporting this study. FL and FT acknowledge the support from the CEA/GENCI (Grand Equipement National de Calcul Intensif) for awarding access to the TGCC (Très Grand Centre de Calcul) Joliot Curie/IRENE supercomputer within the A0110413001 project. RD and EQS are supported by the United States Department of Energy (DOE), Grant No. DE- SC0025420.

Supplemental Information for

# "Collisional energy transfer in ethanimine + He system"

Vivek Vijay, Francesca Tonolo, Ernesto Quintas-Sánchez, Carolin Joy, Richard Dawes, Francois Lique and Dmitri Babikov

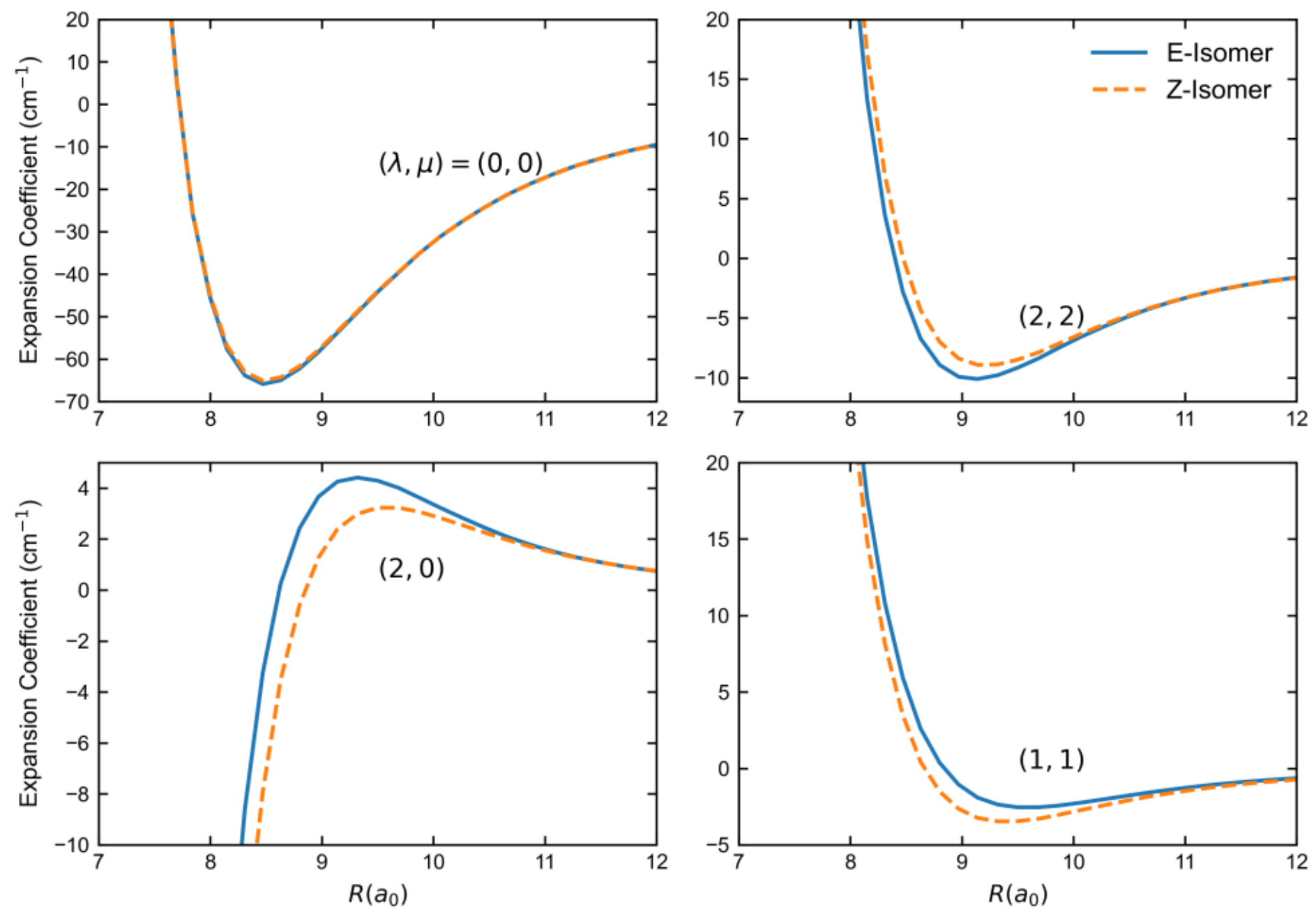

**Figure S1:** Comparison of isotropic and three most important anisotropic expansion coefficients for *E*- and *Z*-isomers of ethanimine.

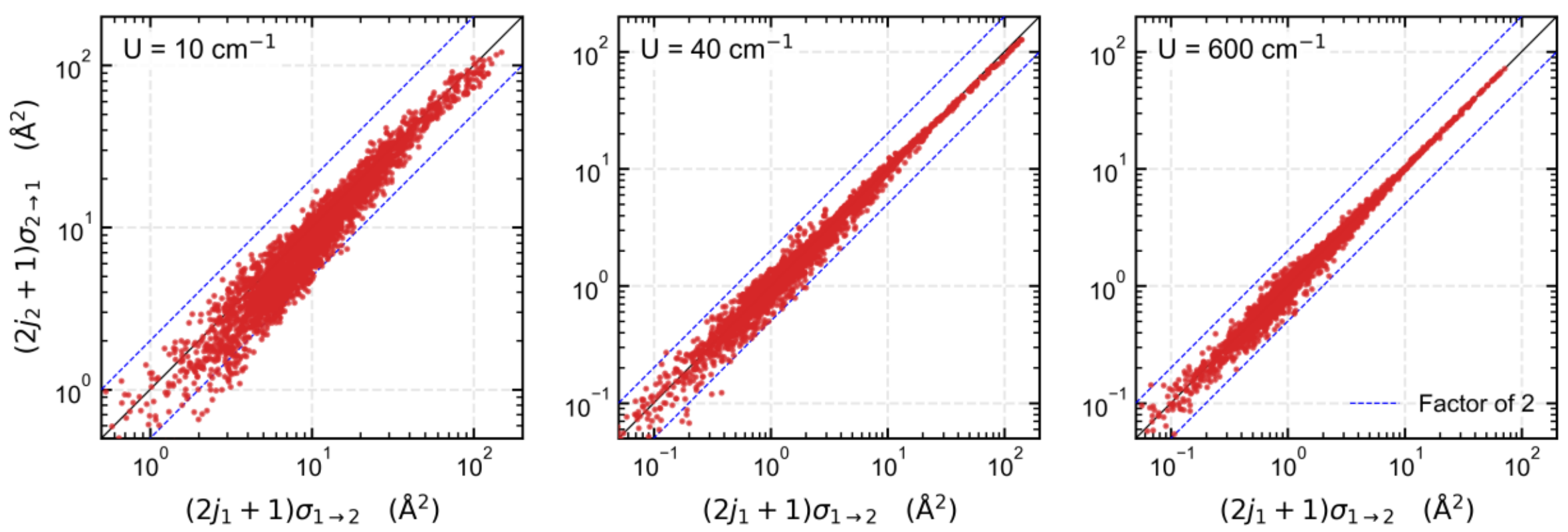

**Figure S2:** A test of MQCT state-to-state transition cross sections with respect to microscopic reversibility at three collision energies. The case of in *E*-isomer of ethanimine + He is presented. The case of *Z*-isomer is very similar. Dashed lines indicate a factor of 2 difference.

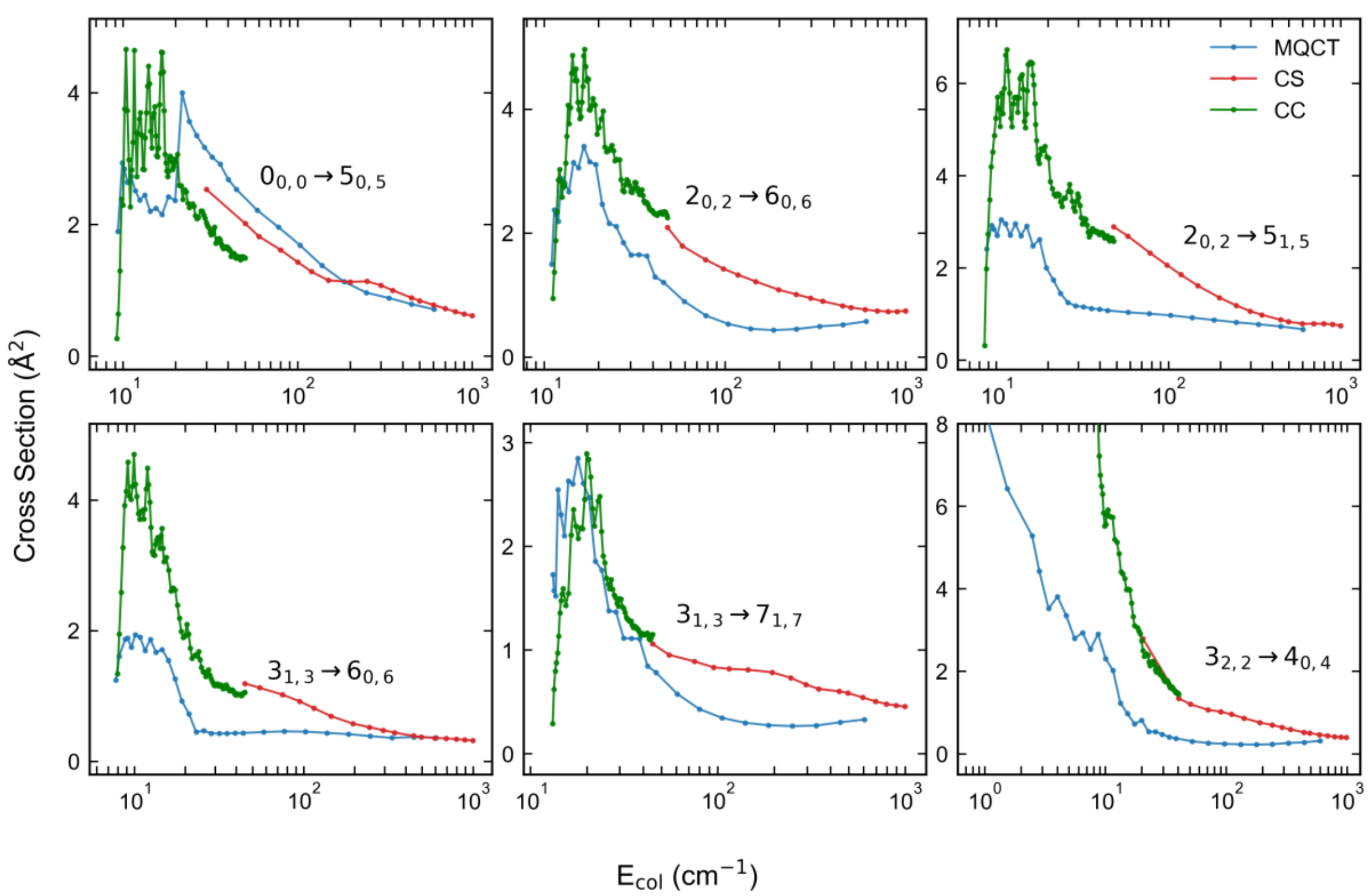

**Figure S3:** Same as Fig. 4 of the main text, but for several weak transitions between the rotational states of *E*-isomer of ethanimine.

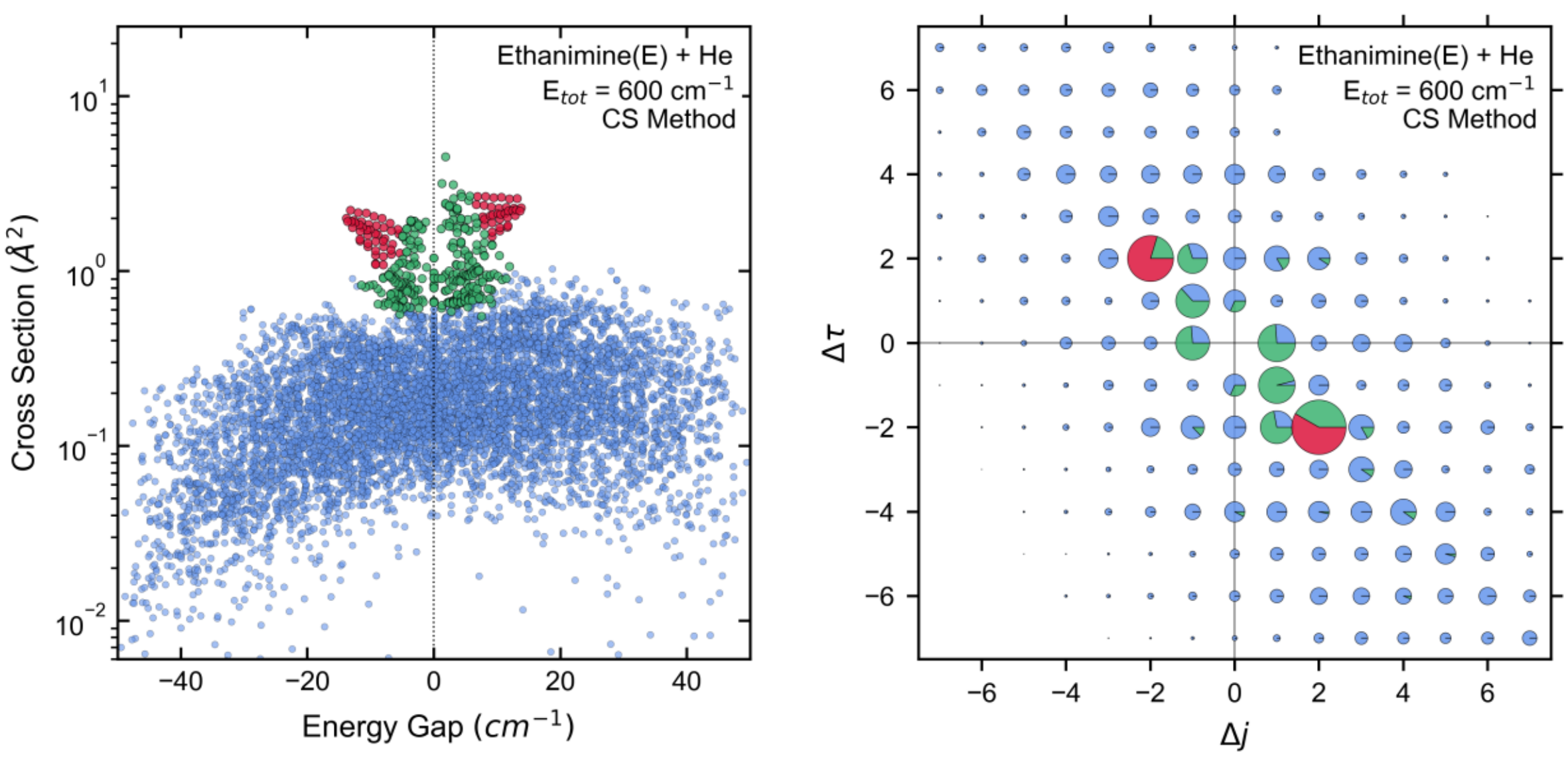

**Figure S4:** Same as Figs. 5 and 6 of the main text but based on cross sections obtained from full-quantum CS calculations at total energy of 600 $cm^{-1}$.

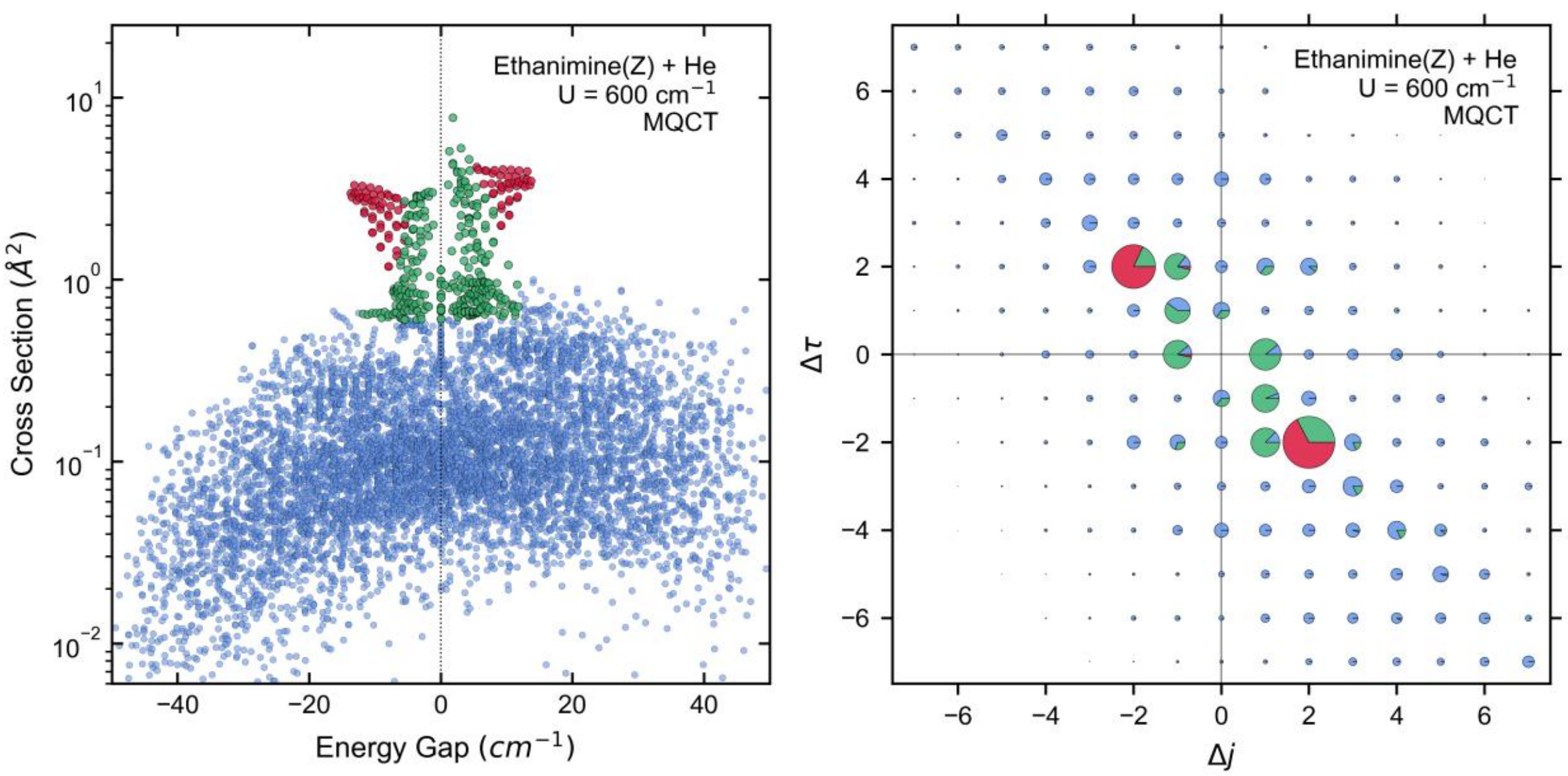

**Figure S5:** Same as Figs. 5 and 6 of the main text but for *Z*-isomer of ethanimine.

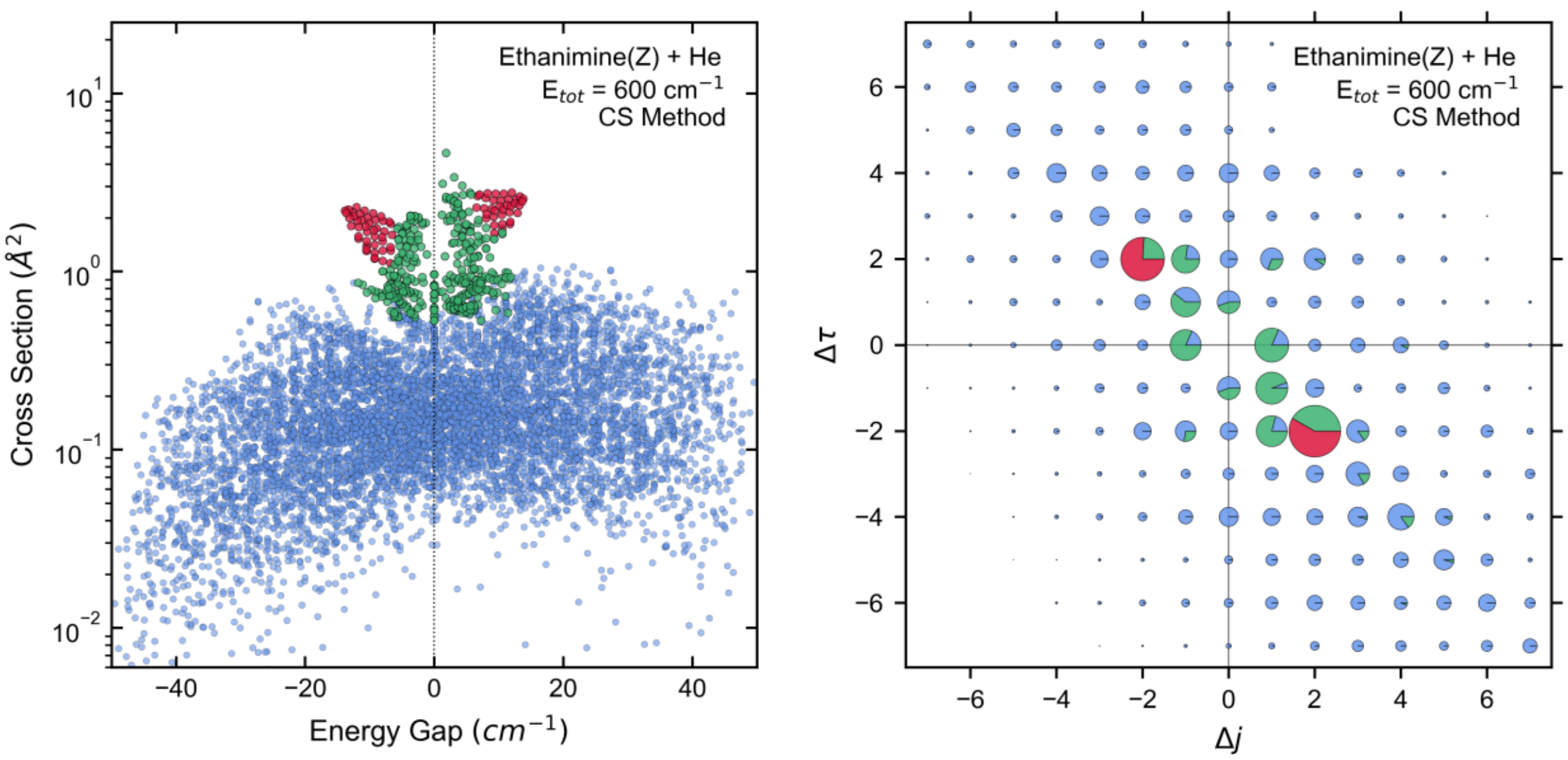

**Figure S6:** Same as Fig. S4 above, but for *Z*-isomer of ethanimine.

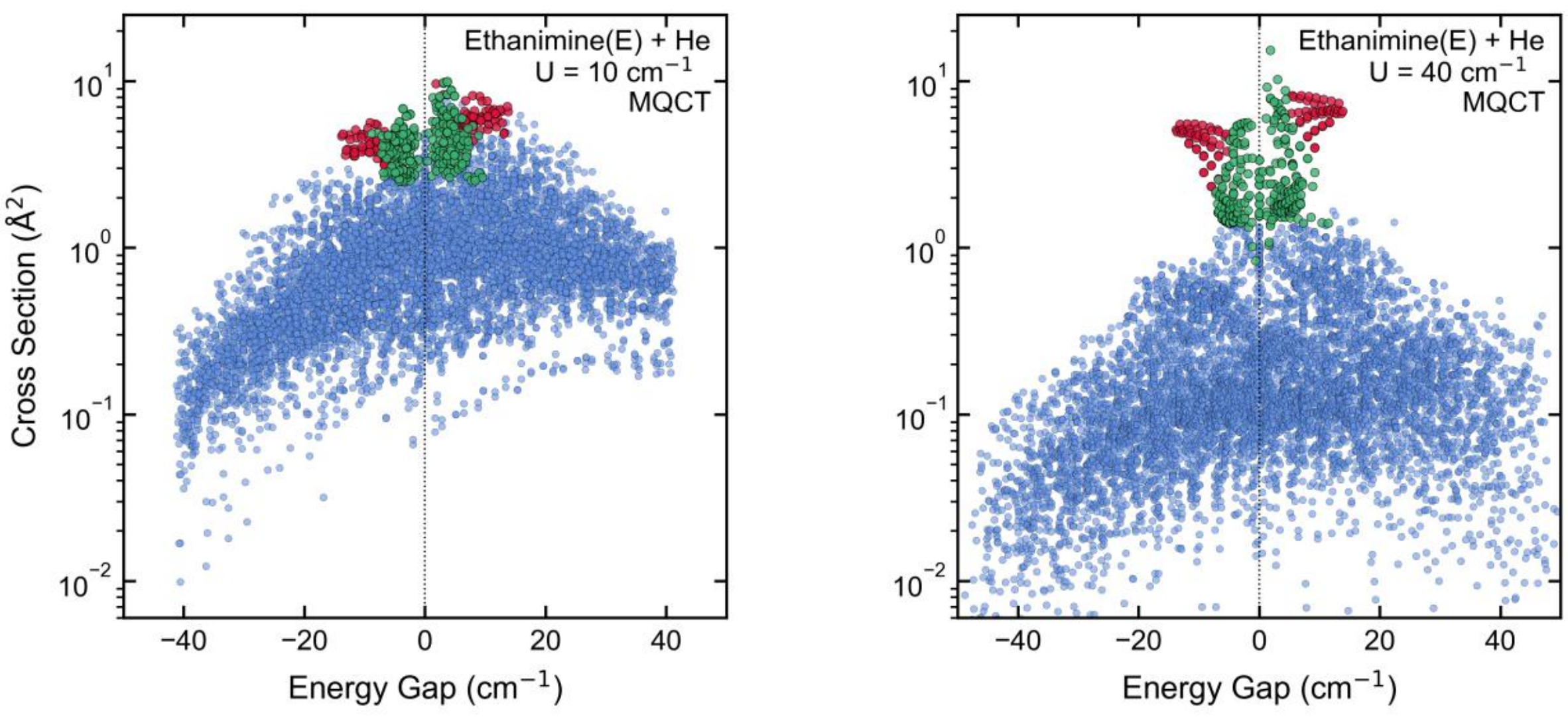

**Figure S7:** Same as Fig. 5 of the main text, but for two values of kinetic energy $U \sim 10\ \mathrm{cm}^{-1}$ and $40\ \mathrm{cm}^{-1}$.

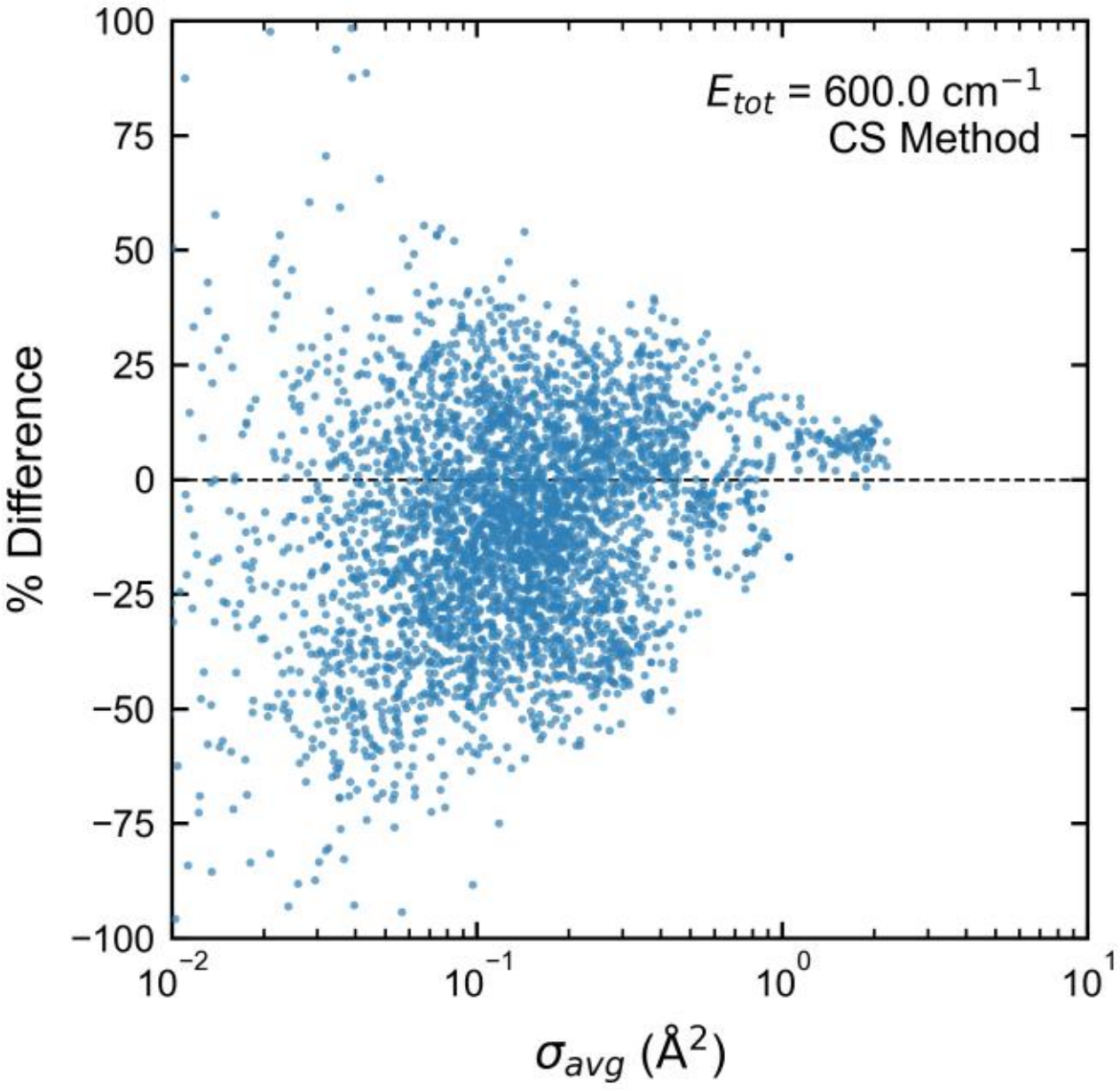

**Figure S8:** Same as Fig. 7 of the main text but based on cross sections obtained from full-quantum CS calculations at total energy of $600\ \mathrm{cm}^{-1}$.